\documentclass[conference]{IEEEtran}
\usepackage[noadjust]{cite}
\usepackage{amsmath,amssymb,amsfonts}
\usepackage{algorithmic}
\usepackage{graphicx}
\usepackage{textcomp}
\usepackage{xcolor}
\usepackage{hyperref}
\usepackage{authblk}
\def\BibTeX{{\rm B\kern-.05em{\sc i\kern-.025em b}\kern-.08em
    T\kern-.1667em\lower.7ex\hbox{E}\kern-.125emX}}
    
\newcommand{\castro}{{\sf Castro}}

\newcommand{\maestroex}{{\sf MAESTROeX}}
\newcommand{\nyx}{{\sf Nyx}}
\newcommand{\amrex}{{\sf AMReX}}
\newcommand{\boxlib}{{\sf BoxLib}}
\newcommand{\microphysics}{{\sf Microphysics}}

\newcommand{\cholla}{{\sf Cholla}}
\newcommand{\gamer}{{\sf GAMER}}
\newcommand{\kathena}{{\sf K-Athena}}
\newcommand{\warpx}{{\sf WarpX}}

\begin{document}

\title{Preparing Nuclear Astrophysics for Exascale}

\author[1]{Max P. Katz}
\author[2]{Ann Almgren}
\author[3]{Maria Barrios Sazo}
\author[3,4]{Kiran Eiden}
\author[2]{Kevin Gott}
\author[3]{Alice Harpole}
\author[2]{\\Jean M. Sexton}
\author[2]{Don E. Willcox}
\author[2]{Weiqun Zhang}
\author[3]{Michael Zingale}
\affil[1]{NVIDIA Corporation}
\affil[2]{Center for Computational Sciences and Engineering, Lawrence Berkeley National Laboratory}
\affil[3]{Department of Physics and Astronomy, Stony Brook University}
\affil[4]{Department of Astronomy, University of California Berkeley}

\maketitle

\begin{abstract}
Astrophysical explosions such as supernovae are fascinating events that require 
sophisticated algorithms and substantial computational power to  model. Castro 
and MAESTROeX are nuclear astrophysics codes that simulate thermonuclear fusion 
in the context of supernovae and X-ray bursts. Examining these nuclear burning 
processes using high resolution simulations is critical for understanding how 
these astrophysical explosions occur. In this paper we describe the changes that 
have been made to these codes to transform them from standard MPI + OpenMP codes 
targeted at petascale CPU-based systems into a form compatible with the pre-exascale 
systems now online and the exascale systems coming soon. We then discuss what new 
science is possible to run on systems such as Summit and Perlmutter that could not 
have been achieved on the previous generation of supercomputers.
\end{abstract}

\begin{IEEEkeywords}
astrophysics, hydrodynamics, computational fluid dynamics, gravity, nuclear reactions,
adaptive mesh refinement
\end{IEEEkeywords}

\section{Introduction}
\label{sec:introduction}

Stellar explosions create the elements that make life possible and are indispensable 
tools for mapping the universe. Many of these events are caused by rapid thermonuclear 
fusion: on time scales much shorter than a second, light elements such as hydrogen and 
helium are fused into heavier elements such as magnesium and iron. The fusion process 
releases copious amounts of energy into a small pocket of space, and the temperature 
of the star is almost instantly raised to billions of degrees. This results in the 
surface of the star being blown out into space (novae and X-ray bursts) or the complete
destruction of the star (supernovae). It is known that thermonuclear explosions happen 
on white dwarf stars, remnants of stars like the Sun, but in many cases it is not known
what triggers these events. Due to the rarity of these events and their great distance 
from the Earth, observational astronomy is not enough to understand their nature and it
must be paired with simulation.

Simulation codes that target these events must be able to simultaneously model the star
as a whole (or at least a large part of it) and the initiation of the explosion process,
which happens much faster and in a much smaller region than the time and length scales 
that define the star as a whole. This calls for some form of adaptivity, where the parts
of the star that may explode are simulated with higher fidelity (and thus more computational
power) than other regions. \maestroex\ (\cite{MAESTRO,MAESTROeX,MAESTROeX-JOSS}) and \castro\ 
(\cite{CastroI}) are astrophysical simulation codes that simulate, respectively, the time 
period leading up to the explosion and the explosion itself. They are built on top of the
\amrex\ (\cite{AMReX-JOSS}) adaptive mesh refinement framework, and provided as part of the
AMReX Astrophysics Suite (\cite{AMReX-Astro}) of simulation codes.

\maestroex\ and \castro\ have been used to study the origins of X-ray bursts 
\cite{MAESTROeX-XRB,Castro-XRB} and thermonuclear\footnote{A thermonuclear supernova, often
called a Type Ia supernova, is the explosion of a white dwarf star as a result of energy release
associated with nuclear fusion. There is another class of supernovae caused by the collapse of
massive stars. We do not examine these core-collapse supernovae in this paper.} supernovae 
\cite{MAESTRO-subchandra,Castro-wdmerger}
on previous generations of supercomputers including Jaguar and Titan at OLCF and Edison 
and Cori at NERSC. MPI was used for parallelism across many nodes, optionally in conjunction 
with OpenMP for intra-node parallelism. These systems enabled simulations that began to
explore the nature of the thermonuclear fusion, but to make the simulations feasible
the simulations were performed at relatively low resolutions and over relatively short
simulation time scales. Unfortunately, thermonuclear explosions are so sensitive that 
low resolution simulations can be qualitatively wrong, and it is clear to us that much 
higher resolution simulations are needed than could have been run on those systems.

Pre-exascale systems that employ accelerators such as Summit and Perlmutter promise to
enable such higher resolution simulations, but a substantial code refactoring effort
was needed to take advantage of these architectures. (The codes had previously been used
on systems such as Blue Waters and Titan but without using their GPUs.) This paper describes 
the software architecture of these astrophysics codes (and their dependencies, including \amrex) 
before and after the transition to GPU-ready status and the process taken to get there. Then 
performance results (at both the single node and many node levels) are presented, and the paper closes
with a discussion of the new science simulation capabilities enabled by GPU-powered supercomputers.

\section{Software Architecture and Development}
\label{sec:software}

\begin{figure*}[htbp]
\centering
\includegraphics[width=0.32\textwidth]{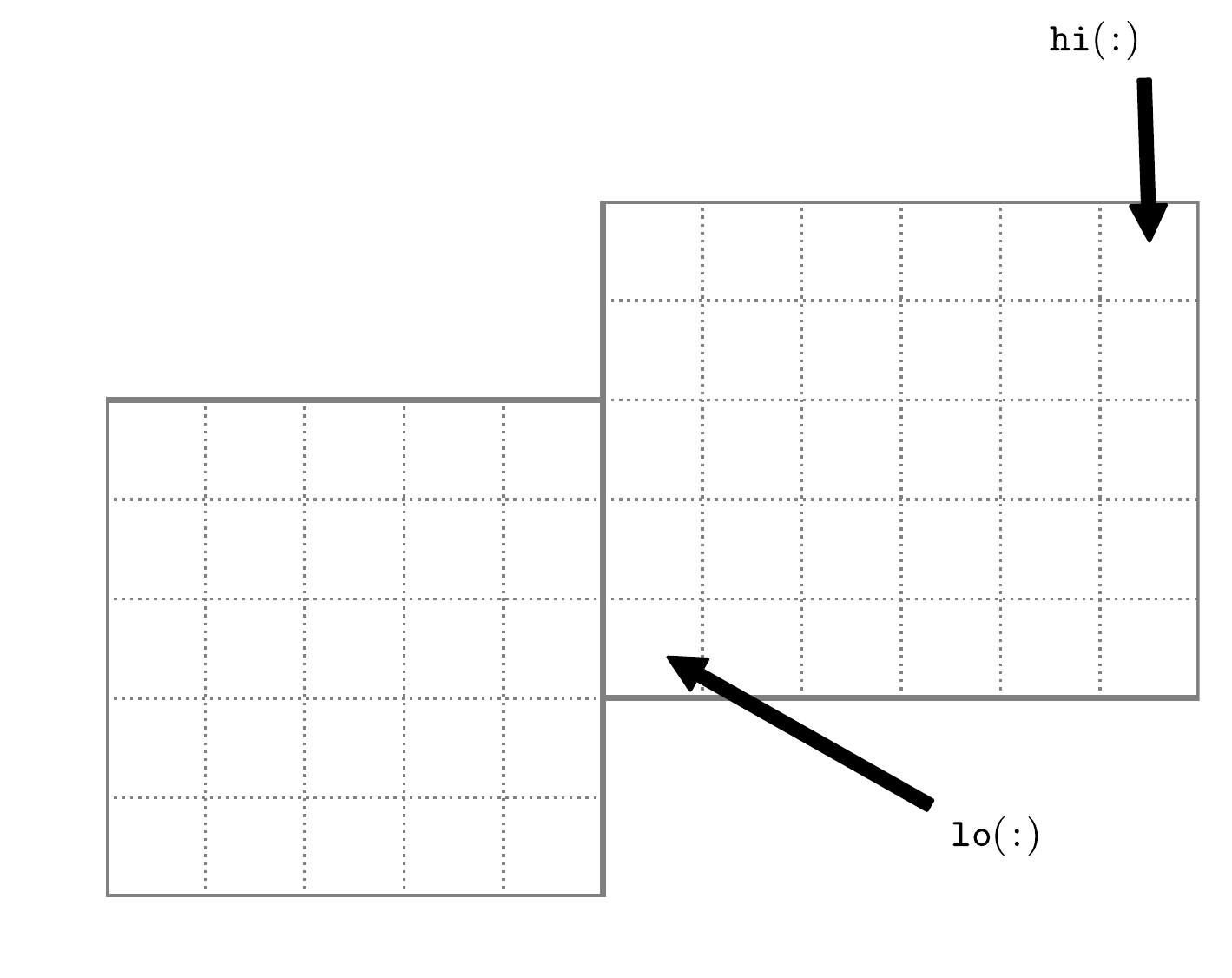}
\includegraphics[width=0.32\textwidth]{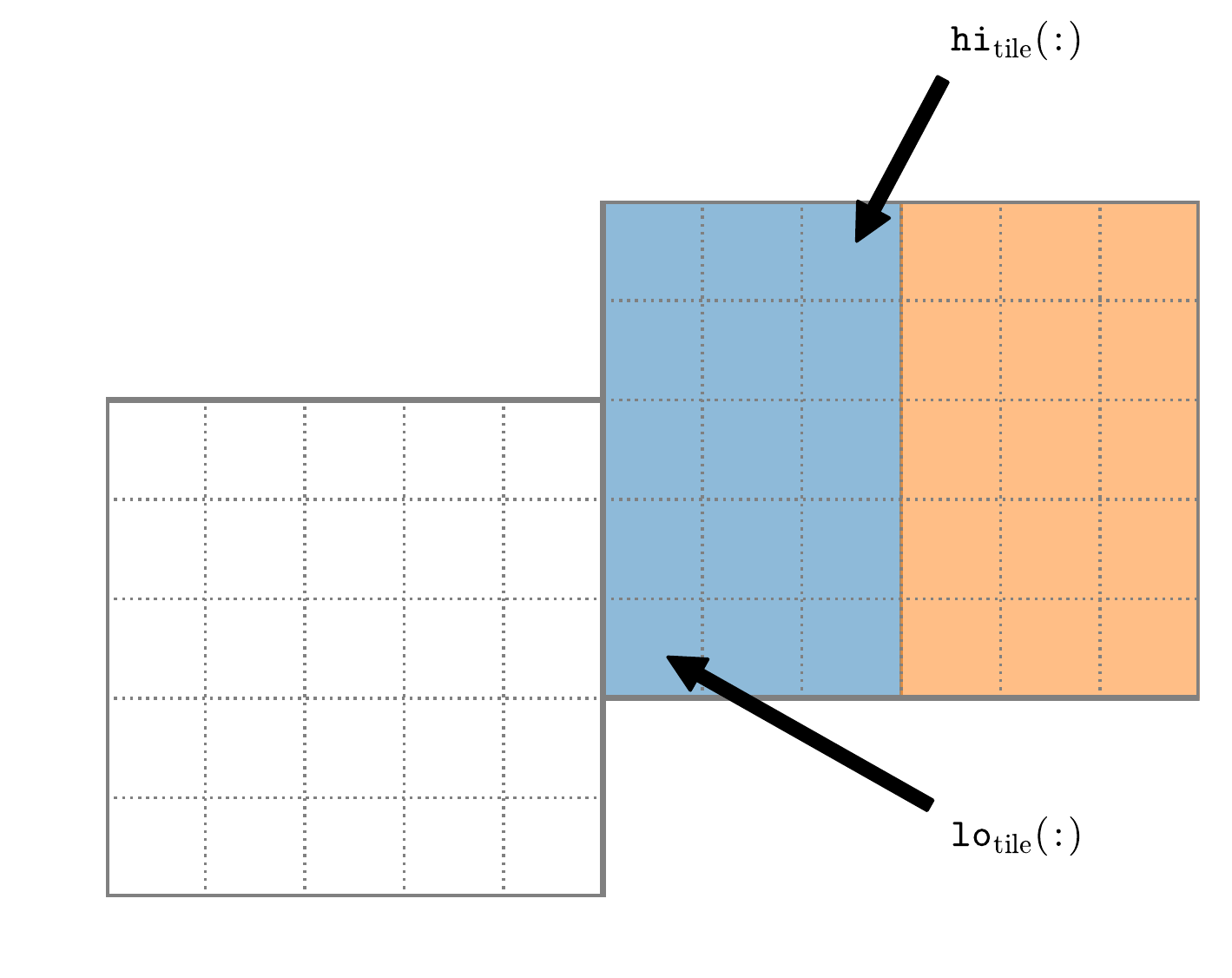}
\includegraphics[width=0.32\textwidth]{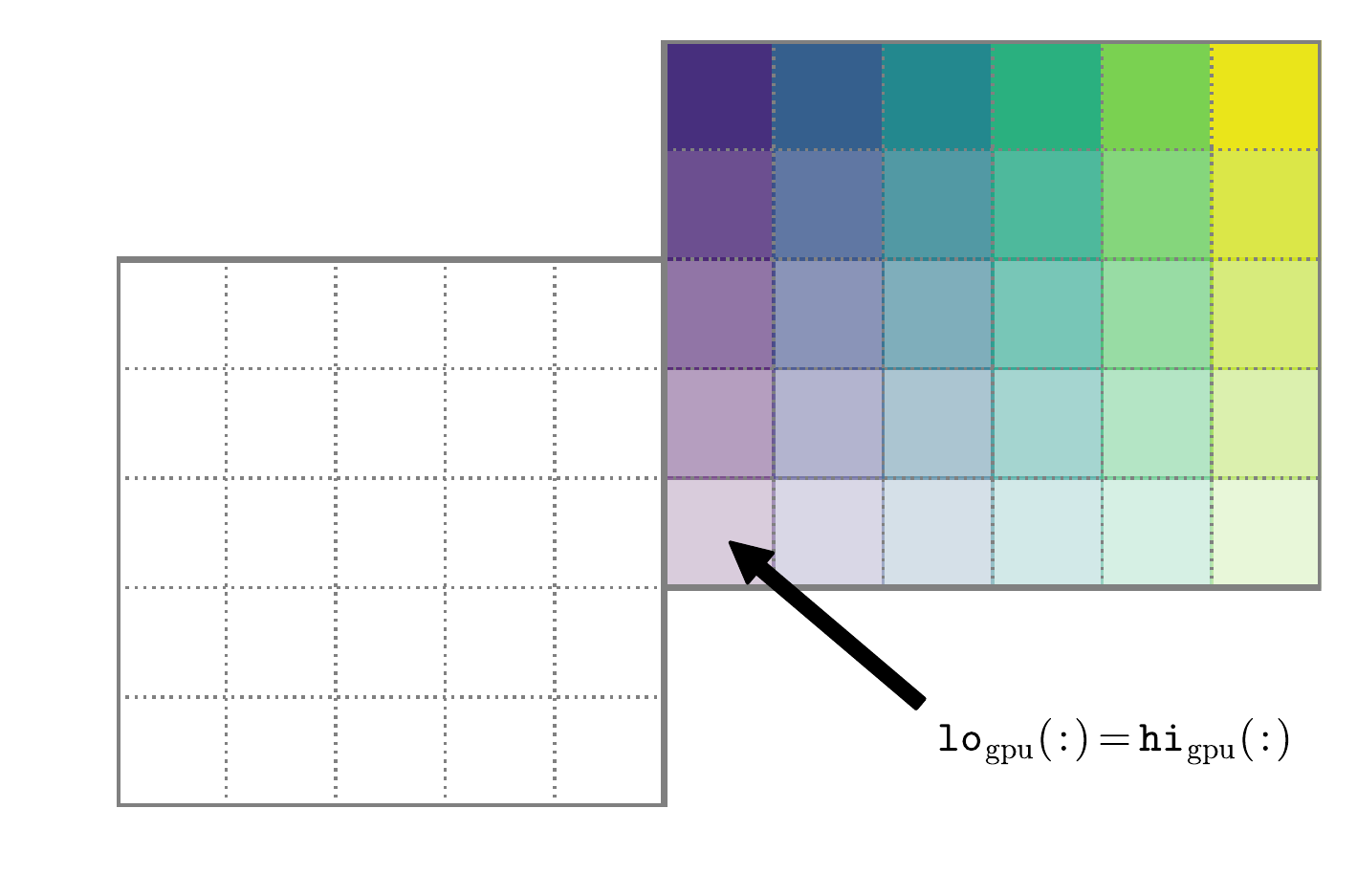}
\caption{(\textit{Left}) A MultiFab lives on a collection of boxes, bounded by
the solid dark lines. Each box (a Fab) is assigned to an MPI rank. The dashed light lines 
denote the zone boundaries. The box runs from the $\mathtt{lo(:)}$ index to the $\mathtt{hi(:)}$
index. (\textit{Center}) A possible division of a box into tiles. An OpenMP thread is assigned to a tile.
(\textit{Right}) On a massively parallel architecture like a GPU, sufficient parallelism is only
exposed when every zone can be threaded; that is, when the $\mathtt{lo}$ and $\mathtt{hi}$ bounds
are the same.}
\label{fig:gridding} 
\end{figure*}

The AMReX Astrophysics Suite of codes includes the grid-based hydrodynamics codes \maestroex, 
\castro, and \nyx\ (\cite{Nyx}). All three rest on top of \amrex, which maintains their gridding capabilities
(including adaptive mesh refinement) and other key driver infrastructure such as memory management.
\amrex's core data concept for grid codes is a MultiFab, which is a (possibly disjoint) set of fluid 
data at a given level of mesh refinement that is distributed across multiple Fabs. (\amrex\ also supports
simulations with particles, which is useful for electromagnetic particle-in-cell codes like \warpx\ 
(\cite{WarpX}) and cosmology codes like \nyx.) A Fab represents
a contiguous section of space, indexed in one, two, or three dimensions, and it can have multiple 
fluid components such as density and temperature (that is, it is a four-dimensional array). \amrex\ 
codes mostly use an MPI + X parallelism model, where MPI is used to distribute Fabs within a MultiFab 
over ranks, and some intra-node parallelism model ("X", where X is a shared memory
parallel programming model like OpenMP, OpenACC, or CUDA) may be used for the Fabs within each rank. Fabs 
are decomposed in a tiled fashion that exposes subsets of the Fab (though the tile may span the whole 
array), and a typical update to the fluid state involves iterating through the full set of tiles among 
all Fabs owned by a rank. On the previous generation of supercomputers such as Cori, coarse-grained OpenMP
was typically employed for intra-node parallelism, where each OpenMP thread worked on a tile of a Fab.
See Figure \ref{fig:gridding} for an example.

The simulation codes themselves implement all physics relevant for the problem setups of interest.
For example, \castro\ includes physics capabilities for hydrodynamics, gravity, thermal diffusion, and
nuclear reactions, and \maestroex\ has a similar set of capabilities. Since there is substantial
overlap in the relevant physics, they both rely on a \microphysics\ repository that provides modules
such as the nuclear reactions. The primary difference between \maestroex\ and \castro\ is that the 
former is designed for (relatively) slowly evolving fluid flow such as convection (and can take very 
large timesteps), while the latter targets rapidly evolving phenonema such as detonations (and takes
correspondingly smaller timesteps). For the problems of interest in this paper, both codes 
feature a combination of locally coupled algorithms (requiring only point-to-point communication
between nearest neighbors) and globally coupled algorithms (requiring all-to-all communication).

Traditionally, codes based on \amrex\ and its predecessor \boxlib\ implemented their physics
kernels in Fortran. A Fortran routine would typically include a triply-nested loop over
\texttt{i}, \texttt{j}, and \texttt{k} ($x$, $y$, and $z$ in space), where the bounds of
the loop are determined by the spatial extent of the tile. Both C++ and Fortran driver interfaces
were exposed; \castro\ and \maestroex\ both use the C++ interface, and the C++ driver would
call the Fortran routine through the standard ISO C binding. The programming model for many \amrex\ 
codes substantially changed from this approach in preparation for the pre-exascale systems, as
discussed in Section \ref{sec:gpu}.

Both \maestroex\ and \castro\ issue new releases monthly, and are co-designed with \amrex\ and \microphysics\
so that a given release version (for example, 20.04) is tagged the same way in all codes, and is compatible 
among all the codes. All of these software packages are freely available in a permissively-licensed
open source form on GitHub. Community contributions in the form of issues and pull requests are permitted,
and Section \ref{sec:gpu} describes a series of contributions we made to these codes to enable them
to run on modern supercomputer architectures.

\section{Porting to GPUs}
\label{sec:gpu}

Our early attempts at preparation of the \amrex\ astrophysics codes for heterogeneous architectures 
focused on programming models that were minimally invasive. In particular, models such as OpenACC
are attractive because they allow incremental porting and allow for the maintenance of a single
source (portable) code base (a property that we deemed important). However, these efforts
were unsuccessful. While the codes primarily used relatively modern Fortran, some components were
borrowed from older, fixed-format Fortran code. The most important example was the ordinary 
differential equation (ODE) integrator, VODE (\cite{VODE}), which contained numerous constructs 
that were not compatible with OpenACC (at least as implemented in the compilers available circa 2015), 
such as computed go to statements. ODE integrators are the key component of nuclear reactions
simulations, and so this approach was not feasible. Even in the cases where code actually compiled,
we did not achieve performance better than available on CPUs. The architecture of the GPUs available 
at the time (e.g. the NVIDIA Kepler K20X GPU on the OLCF Titan system) was extremely unfriendly 
towards the significant amount of thread-local storage needed to evaluate some of the relevant physics.

The first working approach was based on CUDA Fortran. Even for CUDA Fortran, in practice only a 
subset of the language is supported, so a substantial refactoring of the Fortran code was done to
remove all constructs that the PGI compiler effectively disallowed. We still desired portability,
so the choice was made to keep the Fortran subroutines in the same format: a loop over some range
of indices. But now to expose sufficient parallelism for a GPU, each thread has to be assigned a
single zone rather than a tile (Figure \ref{fig:gridding}). This required every part of the algorithm
to be expressed in an embarrassingly parallel manner, with no explicit dependence between zones.

However, some parts of the software had previously been written in a way to maximize expected 
performance on vector CPU architectures. A classic example is the core stencil operation in the 
hydrodynamics routine, where a slope at a given index is calculated by looking at several zones
to the left and right. The CPU code first calculated the slopes for all zones in the $x$ dimension
(say), stored them in a local array, and then read from that array in a second loop that used two
values from the slope array to reconstruct a polynomial approximation of the fluid state. Converting
this to a fully thread parallel format required redundantly calculating two slopes for each zone,
which implied more total floating point operations when summed across all threads, but exposed massive
parallelism. (The original approach can be recovered on NVIDIA GPUs through the use of on-chip
shared memory and synchronization points, which we have not yet explored.) Fortunately, the refactoring
effort ultimately led to a performance improvement on CPUs as well, due largely to decreasing the
memory footprint of the code (the code was memory bandwidth bound on CPUs as well).

This approach was quite promising and early testing on GPU-enabled systems demonstrated a nice speedup for 
pure hydrodynamics tests on the NVIDIA V100 GPU, compared to server-class CPU nodes.
However, a challenge remained: as mentioned previously, the driver framework is in C++, so the way that
GPU offloading was achieved was to write CUDA C++ kernels that called the Fortran routines as device
functions evaluated on a per-zone basis. These Fortran routines themselves often called other Fortran
device functions, sometimes in different Fortran modules. This meant that there was limited capability 
of the compiler to optimize the kernels through inlining and other approaches. Another significant drawback
was that for every array passed from C++ to Fortran, the bounds of the array had to be copied (since in
the Fortran routines the arrays were indexed starting from offsets corresponding to the index within the grid).
For the functions with many Fabs passed to them, simply maintaining the array bounds in per-thread stack
memory resulted in overflowing the 255 available registers on Volta, resulting in register spilling and
very low occupancy. It was not clear how this could be circumvented without a major refactor.

At the same time, many codes running on US Department of Energy systems were evaluating and porting their
codes to run on top of the C++ performance portability frameworks Kokkos (\cite{Kokkos}) and RAJA (\cite{RAJA}).
These frameworks expose abstraction layers where the work done (the body of a loop) is separated from the
implementation of the loop evaluation on a target architecture. The loop body is captured as a lambda function.
\amrex\ implemented a similar approach, tailored to the context of its application base. A generalized ``parallel
for'' construct does the work of generating a CUDA kernel and mapping individual indices to threads; the application
code needs only to define the work done at a given index $(\mathtt{i},\mathtt{j},\mathtt{k})$. The parallel for
construct works correctly for standard CPU builds, and the abstraction layer permits \amrex\ to easily test new
programming model backends without requiring every application developer using the library to learn all 
of the details of the new programming models. 

This approach ended up being quite similar to the ones provided by the previously mentioned performance 
portability frameworks. The reason why \amrex\ provides this as a distinct implementation, rather than requiring
codes to use something like Kokkos, is that many \amrex-based codes, including the ones discussed here, do not use \amrex\ 
merely as a library, but rather as the overall framework for the code, where \amrex\ defines much of the base infrastructure 
and provides a substantial amount of tooling. As a result, \amrex\ has traditionally tended to develop software targeted at 
the most common use cases of its users. For example, the vast majority of parallel loops in \castro\ and \maestroex\ are tightly
nested three-dimensional loops over a contiguous region of logical index space constituting a box (see Fig. \ref{fig:gridding}).
The \amrex\ parallel for construct accepts as arguments this C++ \texttt{Box} object, as well as the lambda defining the work,
making it extremely convenient for application codes to offload loops in a way that is readable in the context of \amrex-based
software. This flexibility permits experimentation with the backend layer without changing user code. We have
tested Kokkos during this effort and found that it could achieve performance comparable to the \amrex\ implementation
of many of our loops (though it typically required some parameter tuning relative to the base choice of CUDA threadblock 
size in Kokkos).

Given the confluence of these and other factors, we began to contribute a major refactoring effort of most of the 
physics code in \castro\ and \maestroex\ from Fortran to C++. In almost all cases, kernels sped up by one or two 
orders of magnitude, resulting in a very large total performance improvement. Our C++ results have been promising 
enough that all future code development we contribute to the AMReX Astrophysics Suite is likely to take place in C++. 
(These results should not be taken
to imply that Fortran is a poor choice for accelerated systems, as we had good success with Fortran.
It was the haphazard combination of C++ and Fortran that was suboptimal for us.)

From a memory management perspective, our design goal from the beginning was to implement all relevant
physics on the accelerators, so that data motion back and forth to the CPU would occur negligibly often.
(This is not possible when the GPU memory is oversubscribed; the code should almost always be run in a
way where this does not happen to obtain optimal performance.) Consequently, we chose CUDA Unified Memory 
due to its ease of use. This strategy was successful and in the present \maestroex\ and \castro\ codes data 
motion between the CPU and GPU only occurs in a few places (checkpointing the simulation state to disk, and 
MPI transfers, depending on whether CUDA-aware MPI is used). The only serious difficulty was that the codes 
frequently allocate temporary memory in the simulation timestep loop for scratch calculations. These frequent 
allocations are tolerable on CPUs but disastrous in CUDA, where memory allocation is orders of magnitude slower. 
Fortunately, \amrex\ had previously implemented a simple  caching (pool) allocator, where in the asymptotic limit 
memory allocations and frees do not actually allocate or free memory on the GPU; rather, they obtain or release 
handles to previously allocated data. Similar approaches are frequently in use elsewhere in the GPU community, 
for example in the Umpire library (\cite{Umpire}). (Major deep learning frameworks such as TensorFlow and PyTorch 
also implement caching allocators for the same reason.) We contributed a change to \amrex\ to make the caching 
allocator the default for the CUDA-enabled build and that resolved the performance obstacle.

Looking forward, we believe we are well situated to take advantage of new computational architectures
as they come online. We have a single source code that can run on both traditional CPU node architectures and 
modern heterogeneous architectures (and in fact our GPU porting work ended up resulting in a net increase in performance 
on CPUs). The \amrex\ abstraction layer for parallel loops allows offload to multiple backends that support lambda functions, 
so at present we can simply run them serially for CPUs (utilizing OpenMP for coarse-grained parallelism) and in parallel on
GPUs. As of the time of writing, \amrex\ has already begun developing backends for HIP and DPC++, targeting
OLCF Frontier and ALCF Aurora, respectively, and many of the core constructs already work on those architectures and have been
tested on the currently available versions of AMD and Intel hardware. Of course, there will still be some technical differences 
to sort out in how each vendor's implementation of lambda capture interfaces with some of our code constructs. We anticipate
that our overall performance limiters on non-NVIDIA GPU architectures will be similar to those on NVIDIA GPUs, 
namely size and bandwidth of the register file and caches, and to a lesser extent DRAM bandwidth and size.

On the memory front, it is unclear whether those platforms will have an analogue of CUDA's Unified Memory, 
or if so what its performance characteristics would be, but we do not anticipate this being a serious problem. The \amrex\ memory 
allocator wrapper can easily switch between different underlying memory allocation APIs provided by the platform. Because our codes 
are now in the state where the data is already on the GPU for the duration of the simulation, it will not be difficult to switch to 
a memory allocation API that provides device-only data, as we do not ever dereference our simulation data on the CPU. (When we write 
a checkpoint file, it involves making a copy to CPU memory, not migrating the data to the CPU.) We can already prototype this in CUDA, 
and so we will be able to do much of the preparation work for these architectures ahead of time.

\section{Performance Measurements}
\label{sec:performance}

In a hydrodynamics code, performance is typically measured in terms of throughput: how many zones are updated in a given amount
of walltime? A convenient scale is zones updated per microsecond, which has the property that it is O(1) for a modern
high-end CPU server node running a standard pure hydrodynamics test case such as the one presented in Section \ref{sec:sedov}. 
This metric, or the related zones/second metric, has been reported for a few other astrophysical fluid dynamics codes. 
\cholla\ (\cite{Cholla}) reported 7 zones/$\mu$sec for a hydrodynamics algorithm quite similar to the one 
used in \castro\ on the NVIDIA Kepler K20X GPUs on Titan. \gamer\ (\cite{GAMER}) and \kathena\ (\cite{K-Athena}), which
evolve magnetic fields in addition to standard hydrodynamics, report peak throughputs of 55 zones/$\mu$sec on NVIDIA P100 
and 100 zones/$\mu$sec on NVIDIA V100, respectively. Under optimal conditions \castro\ achieves around 25 zones/$\mu$sec
on NVIDIA V100 (this number should not be directly compared against the numbers from other codes, which have different 
algorithms). In this section we describe two test cases that demonstrate our performance with an eye toward scaling behavior. Simulations were run on the OLCF Summit system (IBM AC922 server nodes with 2 IBM Power9 CPUs and 6 NVIDIA V100 GPUs each).

\subsection{Blast Wave}
\label{sec:sedov}

The Sedov-Taylor blast wave (\cite{Taylor1950}) is a standard performance benchmark in computational fluid dynamics,
often implemented in hydrodynamics benchmark codes such as LULESH (\cite{LULESH}) and Laghos (\cite{Laghos}). A source 
of energy is deposited in a very small space at the center of the computational domain; a blast wave then forms
which evolves self-similarly in space and time. In a multiphysics code such as \castro, the Sedov problem exercises
the coupling between the hydrodynamics and equation of state physics modules. The communication pattern for this problem
is local (the hydrodynamics algorithm in \castro\ only requires communication between neighboring regions of the domain.) 
However, the amount of work per zone is relatively small (compared to the science problems of interest for \castro) and 
this problem quickly exposes weak scaling limits as the number of nodes increases.

\begin{figure}[htbp]
\centering
\includegraphics[width=0.48\textwidth]{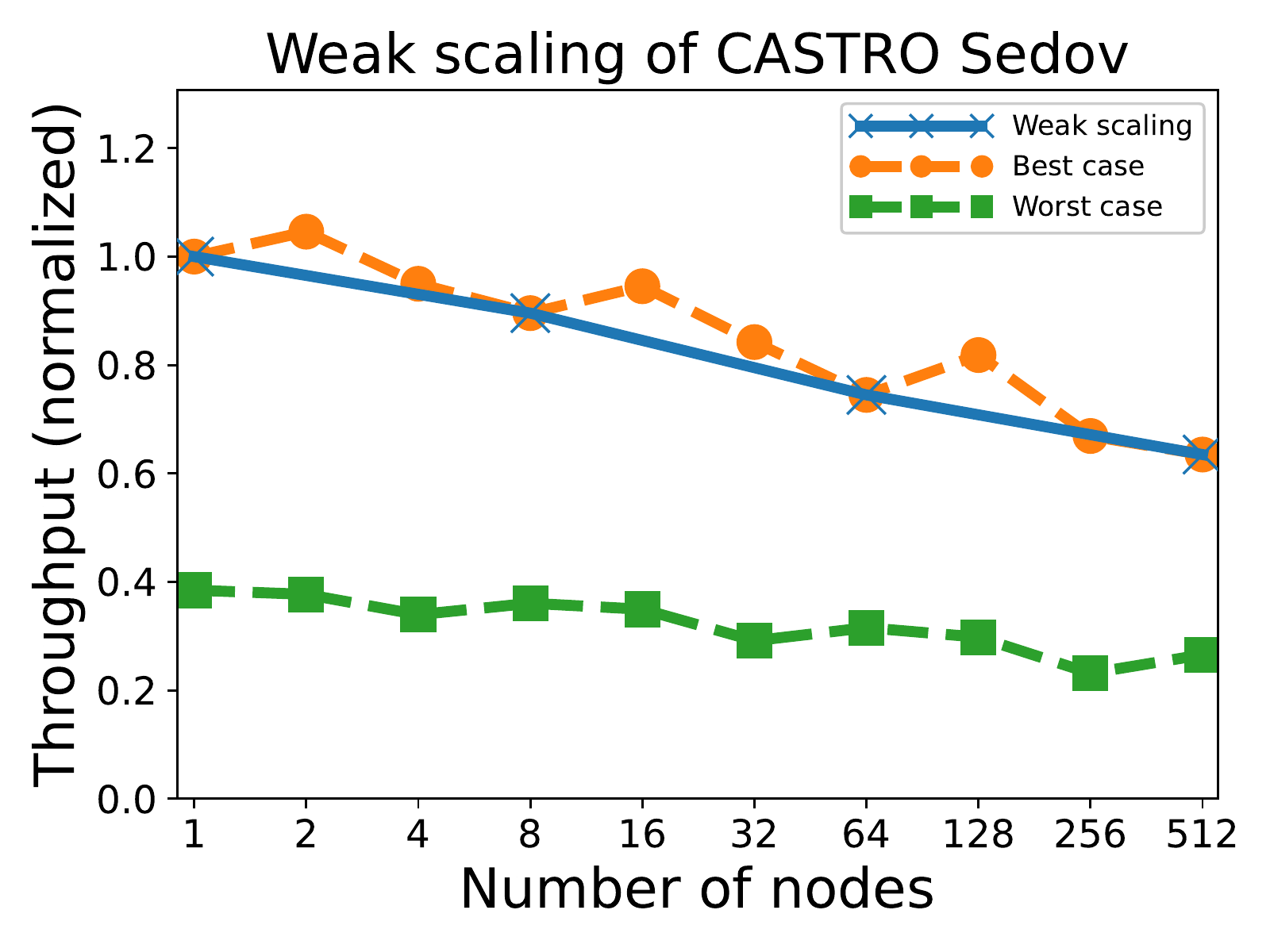}
\caption{\castro\ weak scaling results for the Sedov-Taylor blast wave problem. The horizontal axis is the number of
nodes used, and the vertical axis measures the throughput of the code (number of zones advanced per microsecond),
normalized by dividing the throughput by the single-node throughput and the number of nodes (so that ideal weak
scaling implies a normalized throughput of 1 at all node counts). The solid curve with crosses (the symbols mark the
empirically measured data points, and the line is provided as a fit to guide the eye) is the result for canonical weak scaling: 
the number of nodes is increased by a factor of 8 but the 
work per GPU is kept fixed. The dashed curve with circles is an empirical measurement (again with symbols marking the
measured data points, and a line fit) reflecting a 
notional best case scaling and the dashed curve with squares is a notional worst case 
(the meaning is explained in the text). The single-node throughput for the canonical weak scaling case is 130 zones/$\mu$sec. 
Absolute throughputs for any data point can be reconstructed by multiplying this reference throughput by the number of 
nodes and the normalized throughput.}
\label{fig:sedov}
\end{figure}

Figure \ref{fig:sedov} demonstrates the weak scaling of \castro's implementation of the Sedov-Taylor problem on Summit.
Three scaling scenarios are presented. Weak scaling is typically defined as keeping work per processor fixed while
increasing the number of processors, and for this canonical weak scaling case we distributed a grid of $256^3$ zones across
the 6 GPUs on the node. This test was performed on 1, 8, 64, and 512 nodes, and the results are depicted by the solid 
curve in the figure, with crosses at each data point. The loss in efficiency at higher scales occurs because the code is 
spending more time in message passing (as MPI ranks send their data to other MPI ranks that neighbor them in the simulation domain). 
For reference, the number of zones advanced per $\mu$sec is 130 for one node and approximately 42000 at 512 nodes 
(corresponding to a weak scaling efficiency of about 63\%, the loss in efficiency primarily being due 
to MPI communication overhead).

Many grid codes, including the ones based on \amrex, store the data in blocks rather than as individual data points 
(see the discussion in Section \ref{sec:software}). This means that we cannot arbitrarily divide work among processors:
we decompose the simulation domain in boxes and distribute the boxes among processors. For this test case we distributed
the work in boxes of size $64^3$. This presents a load balancing problem if the number of MPI ranks does not evenly divide
the number of boxes. \castro\ and \maestroex\ use the strategy of one MPI rank per GPU, so there are 6 MPI ranks per node
in this case, which does not evenly divide the 64 boxes in the domain decomposition, so this fiducial case is not an optimal
one for GPU performance. (This is less of a problem for CPU implementations with dozens of MPI ranks per node.) 

There are two other peculiarities associated with running on GPUs. First, GPUs achieve optimal performance by hiding the
latency of individual operations with massive parallelism, so small workloads are inefficient: this discourages very small
boxes (the codes do take advantage of multiple CUDA streams to work on multiple boxes simultaneously, but this only partially
mitigates the effect). Second, GPUs have much smaller memory capacities than CPUs: this discourages very large boxes. 
When the dataset size is larger than the GPU's memory capacity (sometimes called oversubscription), CUDA
Unified Memory can automatically handle this case by evicting some data to make room for other data. However in practice the
performance of this case is currently quite poor on Summit, achieving much lower effective bandwidth during the eviction and replacement
case than the CPU-GPU peak bandwidth (see \cite{Rabbi2020} for an example). We have not attempted to optimize in an attempt to mitigate 
this effect, for example by using the Unified Memory prefetching API to manually move data between the CPU and GPU, and even if we did, 
it is unclear how much performance we could recover. Because the range of box sizes that can meaningfully
fit inside a GPU is limited (an increase in spatial resolution by a factor of 2 increases the amount of data by a factor of 
8, and the problem size that saturates the GPU's compute capacity, $\sim 100^3$ zones, already uses more than 1/8 of the
available GPU memory for most of our setups), we do not attempt a traditional strong scaling measurement. 
It would say much more about the characteristics of the architecture than about how closely we achieved optimality
in our implementation of the physics algorithms, as for small problem sizes we would mostly be measuring the GPU's memory latency,
and for large problem sizes we would mostly be measuring the bandwidth of CPU-GPU transfers during an oversubscription scenario. In practice,
there is a relatively narrow range of grid sizes that make sense to run on GPUs.

Considering these three effects, determining optimal GPU performance and load balancing is a problem that is best solved 
empirically by varying the size of the computational domain and the size of the boxes within it. \amrex\ provides runtime 
parameters to control both the number of zones in the domain, and the  maximum and minimum (per-dimension) widths of a box. 
The domain decomposition algorithm in \amrex\ tends to prefer larger boxes, so for a given domain size, increasing the maximum 
box width effectively decreases the number of boxes. We ran the Sedov test at power-of-two node counts up to 512 nodes, and
in each case we ran several simulations, keeping the minimum box width at 32, but varying the maximum box widths between 32, 
48, 64, 96, and 128. We ran these at two different domain sizes (the smaller domain being a factor of 0.75 smaller per dimension 
than the larger one, e.g. $192^3$ and $256^3$ for the single-node case). The dashed curves in Figure \ref{fig:sedov} represent
the best (circles) and worst (squares) throughput over all of these combinations at each node count (normalized to the value of
the canonical weak scaling throughput on one node). The ``best case'' demonstrates what is possible to achieve with careful tuning by 
the user, while the ``worst case'' demonstrates what happens if the user is careless and neglects to consider these properties.

Our conclusion is that very high performance can be achieved on pre-exascale systems compared to previous architectures, but
careful tuning of runtime parameters is necessary for the user to achieve optimal performance, much more so than was required 
for users of these codes on previous architectures. This is in line with the scaling results of other grid codes (see \cite{GAMER} 
for an example of an order of magnitude performance variation between small and large grids). Unfortunately, this sometimes
encourages the user to select spatial resolutions that are optimal for the system architecture rather than optimal for the
science problem, but that seems to be the price we must pay for accelerated computing on today's architectures.

\subsection{Reacting Bubble}
\label{sec:reacting_bubble}

\maestroex's hydrodynamics algorithm is different from \castro's: the updates require a global communication step
(in particular, a linear solve implemented using the multigrid method). This exposes a fair amount of work but also
significantly more MPI communication in relative terms. However, the core nuclear astrophysics problems of interest
for \maestroex\ involve thermonuclear fusion. The numerical solution of the equations representing the nuclear 
fusion process is purely zone-local work, and takes the form of a coupled set of ODEs. The size of this ODE system 
depends on the number of atomic isotopes represented: a greater number of isotopes generally yields higher fidelity 
but at greater computational expense.

Thermonuclear fusion is a highly energetic process. For many nuclear fusion processes, the energy release is a very
strong function of the temperature $T$ of the surroundings. For example, the energy generation rate of the process that
fuses three helium nuclei into a carbon nucleus may have a temperature dependence as sensitive as $T^{40}$. Other
fusion processes are comparatively less sensitive to the temperature, some dramatically so. As a result there is a wide
range of evolutionary timescales involved, some as short as picoseconds and some much longer. The resulting coupled ODE
system generally needs to use an implicit time integration method (otherwise the whole system would be forced to march
along at the smallest timescale, which would be inefficient). Implicit methods typically require a linear system solve. 
If there are $N$ types of atomic nuclei represented, the size of the matrix representing the linear algebraic system 
is approximately $N^2$, so the computational expense grows quickly with the number of atomic isotopes simulated.

In addition to the compute demands the memory demands are also non-trivial. Modern GPUs such as V100 have a very small
amount of thread-local memory (255 single-precision data registers). With $N \sim 10$ isotopes the Jacobian
of the system alone is enough to fill up these registers, and other physics involved guarantees that we will encounter
register spilling. We are exploring methods to address this, including using sparse linear algebra (many nuclei do not
fuse with other nuclei and the reaction rate is zero, so we do not have to explicitly represent it). But as it
stands now, even though this is a compute intensive process it is not perfectly efficient on GPUs.

\begin{figure}[htbp]
\centering
\includegraphics[width=0.48\textwidth]{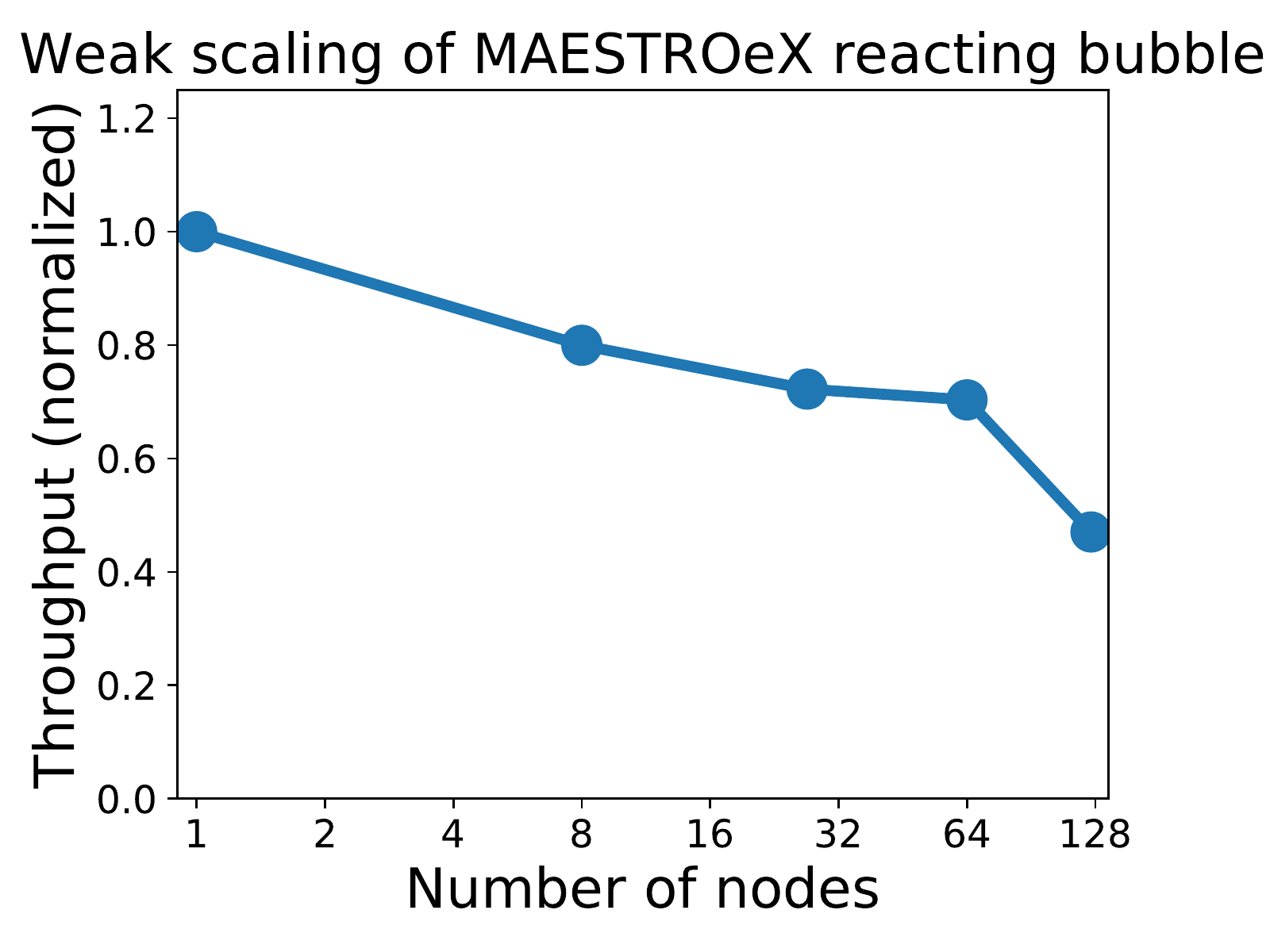}
\caption{\maestroex\ weak scaling results for the reacting bubble problem. The horizontal axis is the number of
nodes used, and the vertical axis measures the throughput of the code (number of zones advanced per microsecond),
normalized to the single-node throughput. The single-node throughput is 11 zones/$\mu$sec. Absolute throughputs 
for any data point can be reconstructed by multiplying this reference throughput by the number of nodes and the 
normalized throughput. Filled circles show the empirically measured data points, and a line fit
is provided to guide the eye.}
\label{fig:reacting_bubble}
\end{figure}

A useful test case demonstrating both the hydrodynamics solve and the nuclear reactions solve is a reacting bubble 
problem \cite{Almgren2008}. A hot bubble is placed in a plane parallel atmosphere. The conditions of the atmosphere 
are similar to those found in the center of the white dwarf stars we are interested in, prior to their potential explosion
as supernovae. This temperature perturbation is large enough to begin localized thermonuclear fusion of carbon in the 
bubble, causing it to become buoyant and rise up through the atmosphere. Despite being a much simpler problem than 
would be modeled in a typical science run (we only model $N = 2$ reacting nuclei), it nevertheless includes all of 
the relevant physical processes. The scaling of this problem is shown in Figure \ref{fig:reacting_bubble}. The problem 
was run on one node and then increased by factors of 2, 3, 4, and 5 per dimension (run on 8, 27, 64, and 125 nodes, respectively).

The performance characteristics of this problem are quite different from the one studied in Section \ref{sec:sedov}.
At single node scale the throughput is approximately 11 zones/$\mu$sec (for reference, this is about a factor of 20
higher than the single-node CPU throughput). The wall time is dominated by the nuclear burning and the parallel 
communication needed for the multigrid solve in the hydrodynamical system, which are approximately equally balanced.
However, at high node counts the parallel communication dominates -- at the highest node count studied, about 6x
more time is spent in the multigrid solve than in the nuclear reactions solve, and the multigrid solve is extremely
communication bound. The globally coupled nature of the system means that it is harder to scale this problem to the
full size of a system like Summit -- but it is also a opportunity, because it means that if we solve the problem
more accurately with a more complete set of atomic nuclei, it will be easier to scale.

While scaling tests are useful, what we really care about is what new science we can investigate. Systems such
as Summit can process a given problem much faster than we would have solved the same problem on a previous supercomputer,
which enables broader parameter studies. But in particular the increased power of these systems enables wonderful new 
science capabilities. Let's now turn to an example of a problem that is well matched to a modern accelerated architecture.

\section{Science Results}
\label{sec:science}

\begin{figure*}[htbp]
\centering
\includegraphics[width=\textwidth]{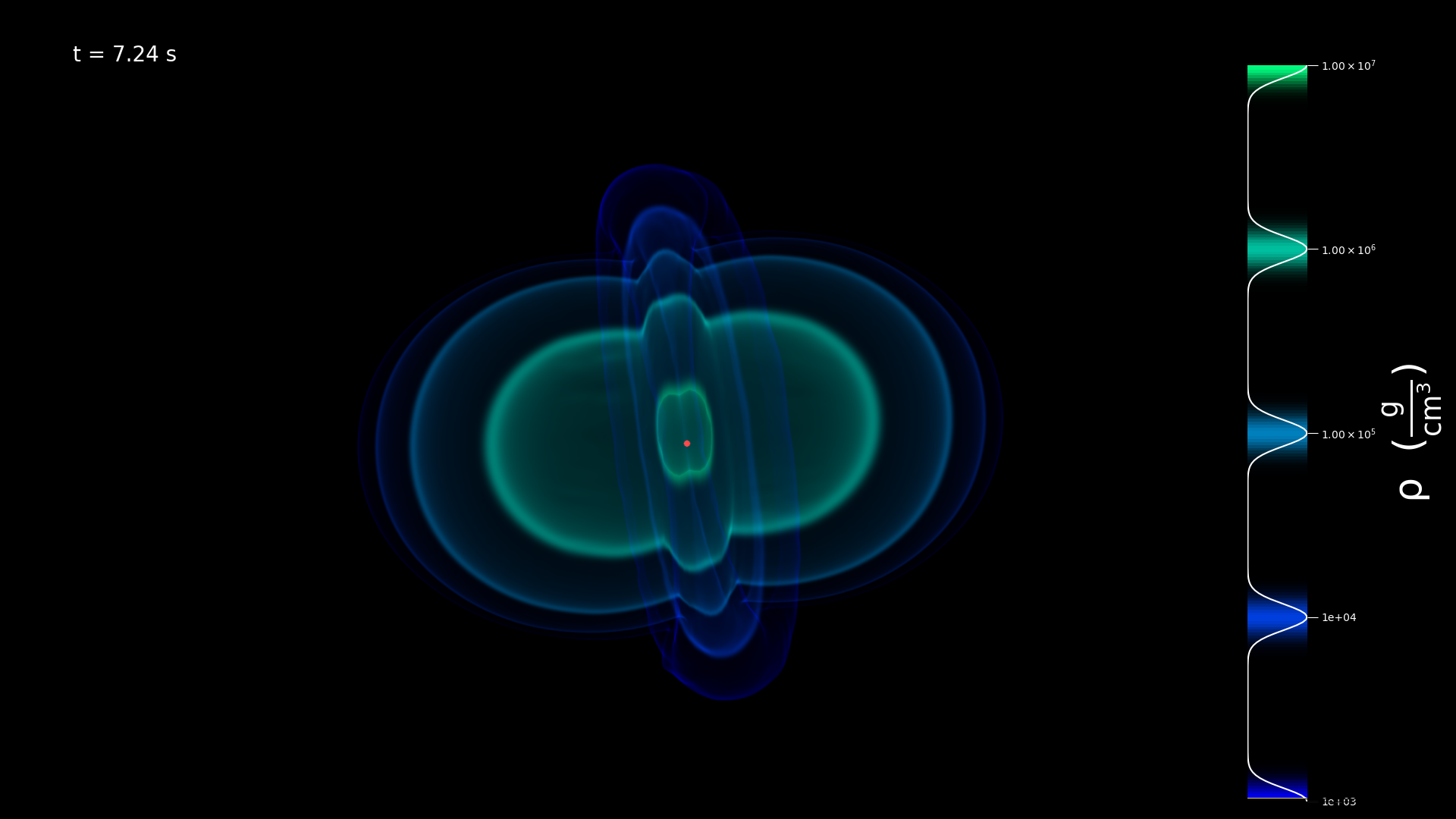}
\\
\vspace{0.1in}
\includegraphics[width=\textwidth]{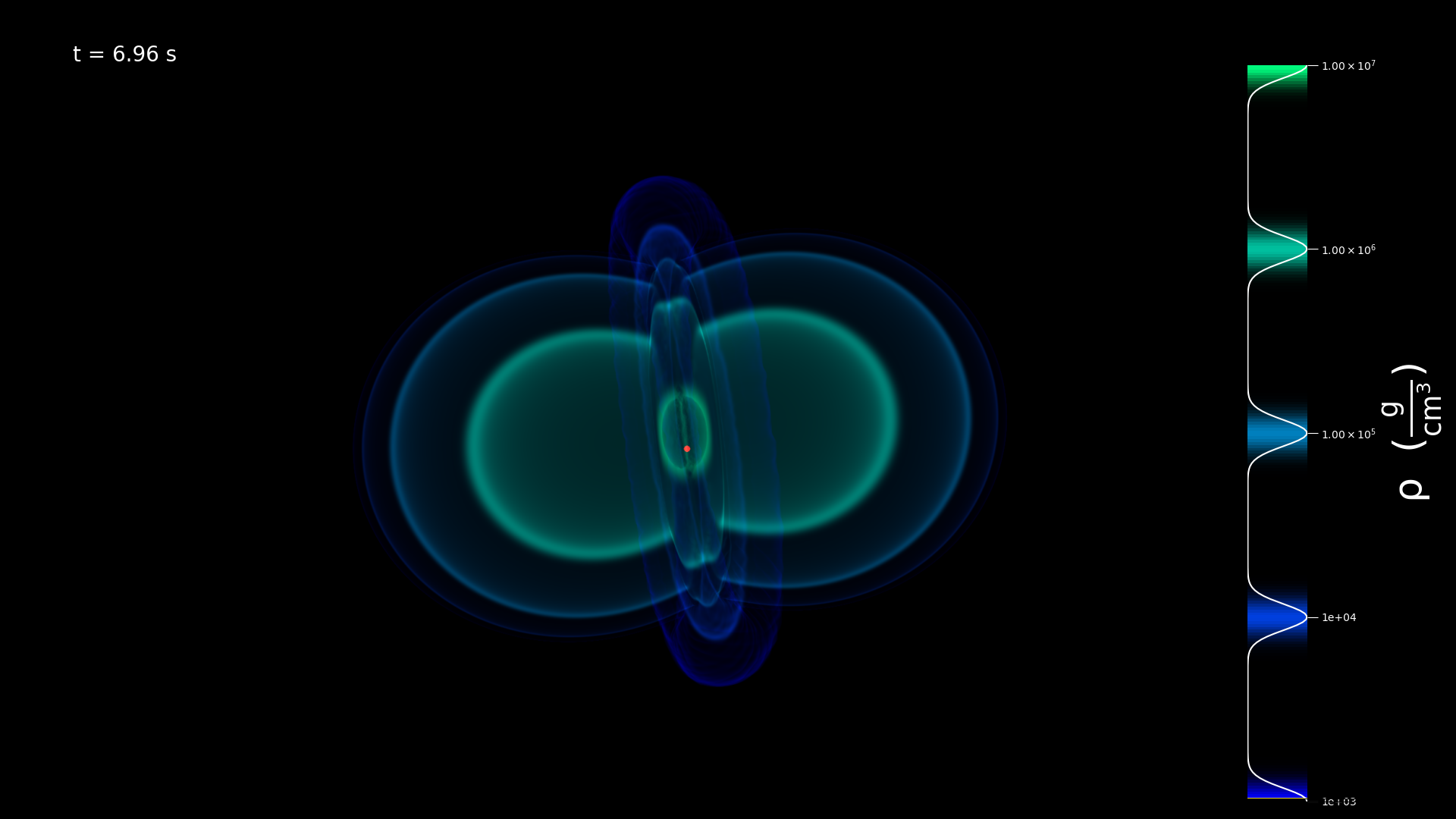}
\caption{Collision of two white dwarfs at the point of thermonuclear ignition (simulation time is noted in the upper left
corner). Contours of the density of the stellar material are shown on a logarithmic scale (scale shown at right). At this time, 
the contact point between the stars is actually denser than the center of either of the stars and has a temperature measured 
in the billions of degrees. The stars are approaching each other from the left and right, respectively, and much of the material 
in the contact point between them has pancaked outward (up and down, from the perspective of the page) since it has nowhere else 
to go. (\textit{Upper}) Low-resolution simulation typical of the state of the art result possible in 2015, which is not numerically converged. A red dot marks the location in space of the thermonuclear ignition, which is at 4 billion Kelvin. (\textit{Lower}) Similar to the upper figure, but we have resolved the contact point by an additional factor of 16 in 
space. Thermonuclear ignition has occurred at an earlier time (though that is partially explained by better resolution of the contact 
point), and the central contact region has still not become very dense. This event would not explain thermonuclear supernovae. However, 
it is also still not fully resolved, so astrophysical conclusions should not be drawn from it.}
\label{fig:collision} 
\end{figure*}

A promising candidate for thermonuclear supernovae is the merging of two white dwarf stars in a binary star system. 
The merger might occur gradually as the stars spiral into each other due to the emission of gravitational waves, as 
predicted by general relativity (\cite{IbenTutukov1984,Webbink1984}). Or it might occur due to the head-on collision of 
two white dwarfs, triggered by oscillations in the binary orbit caused by a nearby companion (\cite{Thompson2011,Hamers2013}).
In this section we examine this collision hypothesis. The collision problem is quite computationally tractable: it is
easy to determine exactly how much simulation time will pass before the stars collide, given an initial setup. When
the stars collide, the kinetic energy of their motion will rapidly be converted into heat, quickly increasing the 
temperature and initiating nuclear reactions. As discussed in Section \ref{sec:reacting_bubble}, these thermonuclear
reactions release copious amounts of energy, heating the nearby material and therefore increasing its reaction rate
in a positive feedback loop. This will almost certainly lead to a thermonuclear detonation, possibly destroying one 
or both stars. The question is only where and when the detonation happens, which determines whether this process is
a possible cause of supernovae or some other astrophysical event. (By comparison, in the gradual inspiral case, it 
is not definitively known if a detonation will occur at any point in the process, so the amount of time needed to 
simulate the system is not well bounded.)

On a system such as Titan at OLCF or Edison at NERSC, we typically would have simulated this process in three dimensions 
on a uniform grid up to $512^3$ in size. We ran such a case on Summit to demonstrate this (the upper panel of Figure \ref{fig:collision}). 
The two white dwarf stars (which are of equal size) are initially situated at a distance equal to twice their diameters, 
moving toward each other at significant speed. The stars collide with each other about seven seconds later (in simulation time) 
and cause rapid thermonuclear fusion. We terminate the run when the temperature reaches four billion Kelvin, as this is the point
at which a detonation and subsequent explosion is imminent. The simplest science question to answer here is: how much time 
passes before the detonation begins? The longer the stars are colliding before the explosion occurs, the more time there 
is to fuse elements at the contact point between the stars into heavier elements such as nickel, the signature of a supernova.
If the detonation occurs promptly when the stars make initial contact, only lighter elements such as magnesium and silicon
will be created, and this event will likely not explain supernovae (but might still be visible as some other astronomical
event).

This problem fits comfortably within 16 nodes and, at that scale, completes in under 15 minutes. This is a significant 
speedup: a typical run on Edison might have used a couple hundred nodes and run on the timescale of many hours. This 
problem is reasonably balanced in performance terms: the early part of the evolution is just the free fall of two objects
toward each other through their mutual gravitational pull. This involves the hydrodynamics module (whose performance was
noted in Section \ref{sec:sedov}) and the gravity module (which involves a global linear solve similar to, though a little
easier than, the one described in Section \ref{sec:reacting_bubble}). The throughput for this part of the evolution is
excellent. When the stars make contact and nuclear fusion begins, the simulation of the nuclear reactions quickly becomes
the bottleneck, becoming about 5 times more expensive than the gravity solve, and the throughput drops substantially. However,
the throughput is still quite good relative to previous systems (nuclear reactions are expensive to simulate on CPUs too),
and the thermonuclear ignition point is soon reached. The fact that we can simulate this whole process in less than ten
node hours means that we could easily now run a parameter sweep over the initial conditions (which would ultimately be 
necessary, given the wide range of possible binary systems in nature) and learn about whether this class of systems might 
explain at least some supernovae.

There is, however, a serious flaw with this type of simulation. Given the size of the domain, the spatial width of a grid 
zone is 50 kilometers (compare to the radius of the stars, nearly 10,000 kilometers, which is the same order of magnitude
as the radius of the Earth -- these stars are quite dense). The length scale over which a thermonuclear detonation begins 
to propagate in this stellar material is not especially well known but is very likely smaller than 10 kilometers 
(\cite{Seitenzahl2009,Garg2017}) and possibly much smaller, so our simulations are unresolved. Worse still, \cite{Kushnir2013} 
and \cite{Katz2019} observe that the thermonuclear fusion process is numerically very unstable at this resolution. 
As we discussed, this type of stellar material is susceptible to thermal feedback loops that rapidly ignite it. (Unlike 
air on Earth, this type of matter does not expand much when heated due to quantum mechanical effects, so the heat from 
nuclear reactions easily gets trapped and causes even more energy release.) Unfortunately, grid zones themselves can 
artificially act as their own cauldrons for this feedback process. This occurs if the timescale for heat to be transferred 
to neighboring zones is much longer than the timescale for heat to be generated in that zone. We inspected the ratio of these 
timescales for simulations at this resolution and found that the energy generation timescale was often an order of magnitude 
smaller than the heat transfer timescale, meaning that we could not distinguish between a physically correct thermonuclear 
detonation and a spurious one caused by numerical instabilities.

As a result, on previous systems we experimented with performing this simulation in two dimensions (using an axisymmetric
cylindrical coordinate system that restricted us to problems symmetric about the axis joining the two stars). This freed up
computational power from the third dimension to be used for adaptive mesh refinement in the contact region between the stars.
In two dimensions, we were finally able to get to the sub-kilometer resolutions necessary to contemplate fully resolving
this process. However, this was only possible for very specific configurations that are physically unrealistic (we cannot
expect the stars to exactly line up when they collide, and it is difficult to simulate processes such as rotation of the
stars about their own axes in an axisymmetric coordinate system).

Now, though, with the computational power unlocked by modern GPU server nodes, we can actually attack this problem in its
full three dimensional glory. The lower panel of Figure \ref{fig:collision} is similar to the upper panel, 
but with higher spatial
resolution. The stars themselves are refined by a factor of 4 at all points in the run (a spatial resolution of 12.5 km).
Additionally, when any material heats up to $10^9$ K, we refine it by an additional factor of 4 (3.125 km). The power of
adaptive mesh refinement is that the additional work added is not proportionally larger (which would be a factor of $4^6$): 
the stars only occupy 0.5\% of the volume of the simulation domain, and the high-temperature region is another factor of
100 smaller in volume. Consequently we can run this setup on only 48 nodes and  still fit it fully in the memory of the 
288 GPUs used. In order to maintain hydrodynamic stability, the length of a timestep is inversely proportional to the 
spatial resolution, so completing the simulation up to the point of ignition still took significantly more time: approximately 
5000 node hours. Nevertheless, this is a new result that qualitatively would not have been possible for us on previous systems.

The key science takeaway from this simulation is that the enhanced resolution helps us answer qualitative questions about the
collision process. The low resolution result ignites at a relatively early time in the collision process: the high density 
contact point has not had enough time to fuse its light elements into heavy elements. This may look like a ``failed'' 
supernova, if we were able to observe it. However, if the ignition had occurred a little bit later, this could have possibly
explained some supernova events, and it was plausible that it would go this way with higher resolution. However, the higher
resolution simulation goes the other way: the ignition occurs even earlier, and there is no way for this process to explain
the supernovae we observe, if we take the simulation result at face value. With that said, we do not believe this simulation 
is numerically converged either, as our two-dimensional simulations suggest we need to achieve at least another order of magnitude 
in spatial  resolution for that, and our diagnostic measurements indicate that the zonal burning timescale is still shorter than 
the energy transfer timescale. While we cannot yet say anything definitive from this result, we now believe we are far closer to a 
realistic answer, and further improvements in computational performance will help us get there.

\section{Conclusion and Outlook}
\label{sec:conclusion}

Stars are three-dimensional objects and need to be modeled that way, with high fidelity required to accurately model
thermonuclear fusion processes. Unfortunately, until now we have been able to achieve either high resolution in two
dimensions, or low resolution in three dimensions, but not high resolution in 3D. Because low resolution simulations are
qualitatively incorrect, we have been pursuing much of our science for the last few years in 2D, for both white dwarf
mergers and X-ray bursts. We have made progress in this way but it is not a path to truly answering our science questions of 
interest.

There is still much work to be done. The simulation of Section \ref{sec:science} was performed by simulating only $N = 13$
elements. This is the bare minimum to say anything qualitatively correct about the thermonuclear fusion process in these
supernovae, and high-fidelity science simulations require at least an order of magnitude more isotopes to be tracked. Furthermore,
we have not achieved the spatial resolution needed to overcome the numerical instabilities associated with simulating thermonuclear 
fusion. We are not sure if it is possible to get a full grasp on this problem on today's systems; however, we are confident we
have made great strides toward that goal.

Our future work will, for the near term, focus primarily on enhancing the performance of the solutions of coupled ODE systems
for the nuclear reactions. There are several avenues opened now that we have ported the ODE integrator (and physics modules that
evaluate the reaction rates) to C++ and substantially modernized it in the process. For example, we can straightforwardly replace
the dense linear system with a sparse linear system. We know what the sparsity pattern is for each combination of isotopes, and that
pattern does not change over time. This allows us to use an optimal sparse representation rather than a generic storage format. Although
using this to decrease memory requirements is most beneficial for very large sets of isotopes, this would be helpful even for the smaller
networks like the one used in this science simulation (which has about 40\% of the dense matrix empty). Along related lines, it is even 
possible to write the exact sequence of operations needed for the linear solve using code generation tools. 

Another avenue of our pursuit relates to using the CPUs as part of the simulation rather than letting them idle. A system like Summit has 
$>95$\% of its peak theoretical computational throughput in GPUs, so one might think that attempting to use the CPUs effectively is fruitless, 
and this is probably true for a problem like our hydrodynamics solve. However, our ODE solves for nuclear reactions have the property that the 
work may be incredibly nonuniform among zones. In the extreme case where one zone in a box is igniting while all of the others are quiescent, the
computational cost may vary by multiple orders of magnitude across zones. This is a challenging task for GPUs, which rely on latency hiding,
because we may not be able to hide the latency when one zone is reacting much longer than the rest. We are currently investigating a strategy
that involves identifying those outlier zones (which will realistically only be a small fraction of the total number of zones) and performing 
their ODE solves on the CPU, while the GPU handles the rest.

With advances in numerical algorithms, the software contributed in this work and related software developed by the HPC community, and 
high-performance accelerated computing architectures, the science problems we really care about are now in reach.

\section*{Acknowledgments}
\castro\ and \maestroex\ are open-source and freely available at \url{https://github.com/AMReX-Astro/}. The work at Stony Brook was supported by DOE/Office of Nuclear Physics grant DE-FG02-87ER40317 and the SciDAC program DOE grant DE-SC0017955. This research was supported by the Exascale Computing Project (17-SC-20-SC), a collaborative effort of the U.S. Department of Energy Office of Science and the National Nuclear Security Administration. MZ acknowledges support from the Simons Foundation. This research used resources of the Oak Ridge Leadership Computing Facility at the Oak Ridge National Laboratory, which is supported by the Office of Science of the U.S. Department of Energy under Contract No. DE-AC05-00OR22725, awarded through the DOE INCITE program. This research has made use of NASA's Astrophysics Data System Bibliographic Services. We also thank NVIDIA Corporation for the donation of a Titan X Pascal and Titan V to Stony Brook University.


\begin{thebibliography}{00}
\bibitem{MAESTRO} Nonaka, A., Almgren, A. S., Bell, J. B., Lijewski, M. J., Malone, C. M., and Zingale, M. (2010). MAESTRO: An Adaptive Low Mach Number Hydrodynamics Algorithm for Stellar Flows. The Astrophysical Journal Supplement Series, 188, 358. doi:10.1088/0067-0049/188/2/358
\bibitem{MAESTROeX} Fan, D., Nonaka, A., Almgren, A. S., Harpole, A., and Zingale, M. (2019). MAESTROeX: A Massively Parallel Low Mach Number Astrophysical Solver. The Astrophysical Journal, 887, 212. doi:10.3847/1538-4357/ab4f75
\bibitem{MAESTROeX-JOSS} Fan, D., Nonaka, A., Almgren, A. S., Harpole, A., and Zingale, M. (2019). MAESTROeX: A Massively Parallel Low Mach Number Astrophysical Solver. Journal of Open Source Software, 4(44), 1757. doi:10.21105/joss.01757
\bibitem{CastroI} Almgren, A. S., Beckner, V. E., Bell, J. B., Day, M. S., Howell, L. H., Joggerst, C. C., Lijewski, M. J., Nonaka, A., Singer, M., and Zingale, M. (2010). CASTRO: A New Compressible Astrophysical Solver. I. Hydrodynamics and Self-Gravity. The Astrophysical Journal 715, 1221. doi:10.1088/0004-637x/715/2/1221
\bibitem{AMReX-JOSS} Zhang, W., Almgren, A., Beckner, V., Bell, J., Blaschke, J., Chan, C., Day, M., Friesen, B., Gott, K., Graves, D., Katz, M., Myers, A., Nguyen, T., Nonaka, A., Rosso, M., Williams, S., and Zingale, M. (2019). AMReX: a framework for block-structured adaptive mesh refinement, Journal of Open Source Software, 4(37), 1370. doi:10.21105/joss.01370
\bibitem{AMReX-Astro} Zingale, M., Almgren, A.S., Barrios Sazo, M. G., Beckner, V. E., Bell, J. B., Friesen, B., Jacobs, A. M., Katz, M. P., Malone, C. M., Nonaka, A. J., Willcox, D. E., and Zhang, W. (2018). Meeting the Challenges of Modeling Astrophysical Thermonuclear Explosions: Castro, Maestro, and the AMReX Astrophysics Suite. Journal of Physics: Conference Series 1031, 012024. doi:10.1088/1742-6596/1031/1/012024
\bibitem{MAESTROeX-XRB}  Zingale, M., Malone, C. M., Nonaka, A., Almgren, A. S., and Bell, J. B. (2015). Comparisons of Two- and Three-Dimensional Convection in Type I X-Ray Bursts. The Astrophysical Journal, 807, 60. doi:10.1088/0004-637X/807/1/60
\bibitem{Castro-XRB} Zingale, M., Eiden, K., Cavecchi, Y., Harpole, A., Bell, J. B., Chang, M., Hawke, I., Katz, M. P., Malone, C. M., Nonaka, A. J., Willcox, D. E., and Zhang, W. (2019).  Toward Resolved Simulations of Burning Fronts in Thermonuclear X-ray Bursts. Journal of Physics: Conference Series, 112, 012005. doi:10.1088/1742-6596/1225/1/012005
\bibitem{MAESTRO-subchandra} Jacobs, A. M., Zingale, M., Nonaka, A., Almgren, A. S., and Bell, J. B. (2016). Low Mach Number Modeling of Convection in Helium Shells on Sub-Chandrasekhar White Dwarfs. II. Bulk Properties of Simple Models. The Astrophysical Journal, 827, 84. doi:10.3847/0004-637X/827/1/84
\bibitem{Castro-wdmerger} Katz, M. P., Zingale, M., Calder, A. C., Swesty, F. D., Almgren, A. S., and Zhang, W. (2016). White Dwarf Mergers on Adaptive Meshes I. Methodology and Code Verification. The Astrophysical Journal, 819, 94. doi:10.3847/0004-637X/819/2/94
\bibitem{Nyx} Almgren, A., Bell, J., Lijewski, M., Lukić, Z., Van Andel, E. (2013). Nyx: A Massively Parallel AMR Code for Computational Cosmology. The Astrophysical Journal, 765, 39. doi:10.1088/0004-637X/765/1/39
\bibitem{WarpX} Vay, J. L., Almgren, A., Bell, J., Ge, L., Grote, D. P., Hogan, M., Kononenko, O., Lehe, R., Myers, A., Ng, C., and Park, J. (2018). Warp-X: A new exascale computing platform for beam–plasma simulations. Nuclear Instruments and Methods in Physics Research Section A: Accelerators, Spectrometers, Detectors and Associated Equipment, 909, 476. doi:10.1016/j.nima.2018.01.035
\bibitem{FLASH} Fryxell, B., Olson, K., Ricker, P., Timmes, F. X., Zingale, M., Lamb, D. Q., MacNeice, P., Rosner, R., Truran, J. W., Tufo, H. (2000). FLASH: An Adaptive Mesh Hydrodynamics Code for Modeling Astrophysical Thermonuclear Flashes. The Astrophysical Journal Supplement Series, 131, 273. doi:10.1086/317361
\bibitem{VODE} Brown, P. N., Byrne, G. D., and Hindmarsh, A. C. (1988). VODE: A Variable-Coefficient ODE Solver. 
SIAM Journal on Scientific and Statistical Computing, 10, 1038. doi:10.1137/0910062
\bibitem{Kokkos} Edwards, H. C., Trott, C. R., and Sunderland, D. (2014). Kokkos: Enabling manycore performance portability through polymorphic memory access patterns. Journal of Parallel and Distributed Computing, 74, 3202. doi:10.1016/j.jpdc.2014.07.003
\bibitem{RAJA} Beckingsale, D. A., Burmark, J., Hornung, R., Jones, H., Killian, W., Kunen, A. J., Pearce, O., Robinson, P., Ryujin, B. S., Scogland, T. R. W. (2019). RAJA: Portable Performance for Large-Scale Scientific Applications. 2019 IEEE/ACM International Workshop on Performance, Portability and Productivity in HPC (P3HPC).
\bibitem{Umpire} Beckingsale, D., Mcfadden, M., Dahm, J., Pankajakshan, R., and Hornung, R. (2019). Umpire: Application-Focused Management and Coordination of Complex Hierarchical Memory. IBM Journal of Research and Development. doi:10.1147/JRD.2019.2954403
\bibitem{Cholla} Schneider, E. E., and Robertson, B. E. (2015). Cholla : A New Massively-Parallel Hydrodynamics Code For Astrophysical Simulation. The Astrophysical Journal Supplement Series, 217, 24. doi:10.1088/0067-0049/217/2/24
\bibitem{GAMER} Zhang, U., Schive, H., and Chiueh, T. (2018). Magnetohydrodynamics with GAMER. The Astrophysical Journal Supplement Series, 236, 50. doi:10.3847/1538-4365/aac49e
\bibitem{K-Athena} Grete, P., Glines, F. W., and O'Shea, B. W. (2019). K-Athena: a performance portable structured grid finite volume magnetohydrodynamics code. arxiv:1905.04341
\bibitem{Taylor1950} Taylor, G. I. (1950). The formation of a blast wave by a very intense explosion I. Theoretical discussion. Proceedings of the Royal Society A, 201, 149. doi:10.1098/rspa.1950.0049
\bibitem{LULESH} Hornung, R. D., Keasler, J. A., and Gokhale, M. B. (2011). Hydrodynamics Challenge Problem. Lawrence Livermore National Laboratory Technical Report, LLNL-TR-490254.
\bibitem{Laghos} Dobrev, V., Kolev, T., and Rieben, R. (2012). High-order curvilinear finite element methods for Lagrangian hydrodynamics. SIAM Journal on Scientific Computing, 34, B606. doi:10.1137/120864672
\bibitem{Rabbi2020} Rabbi, F., Daley, C., Aktulga, H., and Wright, N. (2020). Evaluation of Directive-based GPU Programming Models on a Block Eigensolver with Consideration of Large Sparse Matrices. Lawrence Berkeley National Laboratory. https://escholarship.org/uc/item/83t4f5vj
\bibitem{Almgren2008} Almgren, A. S., Bell, J. B., Nonaka, A. and Zingale, M. (2008). Low Mach number modeling of type Ia supernovae. III. Reactions. The Astrophysical Journal, 684, 449. doi:10.1086/590321
\bibitem{IbenTutukov1984} Iben, I., Jr., and Tutukov, A. V. (1984). Supernovae of type I as end products of the evolution of binaries with components of moderate initial mass (M not greater than about 9 solar masses). The Astrophysical Journal Supplement Series, 54, 335. doi: 10.1086/190932
\bibitem{Webbink1984} Webbink, R. F. (1984). Double white dwarfs as progenitors of R Coronae Borealis stars and Type I supernovae. The Astrophysical Journal, 277, 355. doi:10.1086/161701
\bibitem{Thompson2011} Thompson, T. A. (2011). Accelerating Compact Object Mergers in Triple Systems with the Kozai Resonance: A Mechanism for ``Prompt'' Type Ia Supernovae, Gamma-Ray Bursts, and Other Exotica. The Astrophysical Journal, 741, 82. doi:10.1088/0004-637X/741/2/82
\bibitem{Hamers2013} Hamers, A. S., Pols, O. R., Claeys, J. S. W., and Nelemans, G. (2013). Population synthesis of triple systems in the context of mergers of carbon-oxygen white dwarfs. Monthly Notices of the Royal Astronomical Society, 430, 2262. doi:10.1093/mnras/stt046
\bibitem{Seitenzahl2009} Seitenzahl, I. R., Meakin, C. A., Townsley, D. M., Lamb, D. Q., and Truran, J. W. (2009). Spontaneous Initiation of Detonations in White Dwarf Environments: Determination of Critical Sizes. The Astrophysical Journal, 696, 515. doi:10.1088/0004-637X/696/1/515
\bibitem{Garg2017} Garg, U., and Chang, P. (2017). A Semi-analytic Criterion for the Spontaneous Initiation of Carbon Detonations in White Dwarfs. The Astrophysical Journal, 836, 189. doi:10.3847/1538-4357/aa5d58
\bibitem{Kushnir2013} Kushnir, D., Katz, B., Dong, S., Livne, E., and Fern{\'a}ndez, R. (2013). Head-on Collisions of White Dwarfs in Triple Systems Could Explain Type Ia Supernovae. The Astrophysical Journal Letters, 778, L37. doi:10.1088/2041-8205/778/2/L37
\bibitem{Katz2019} Katz, M. P., and Zingale, M. (2019). Numerical Stability of Detonations in White Dwarf Simulations. The Astrophysical Journal, 874, 169. doi:10.3847/1538-4357/ab0c00
\end{thebibliography}
\end{document}